\documentclass[twocolumn,prl,superscriptaddress,showpacs,byrevtex]{revtex4}
\usepackage{mathptmx,amssymb}
\usepackage{amsmath}
\usepackage{graphicx}
\usepackage{color}
\usepackage{epstopdf}
\usepackage{dcolumn}
\usepackage{bm}
\usepackage[utf8]{inputenc}
\usepackage[francais]{babel}
\DecimalMathComma
\usepackage{amsfonts}
\usepackage{tabularx}
\usepackage{subfigure}

\DeclareTextSymbol{\deg}{T1}{6}
\DeclareTextSymbol{\deg}{OT1}{23}

\usepackage{sistyle}
\usepackage{epsfig, graphicx, xcolor}
\usepackage{subfigure}
\usepackage{tabularx}

\begin{document}
\title[VKM]{1/f noise and long-term memory of coherent structures in a turbulent shear flow}

\author{M. Pereira, C. Gissinger, S. Fauve}
\affiliation{Laboratoire de physique statistique, Ecole normale sup\'erieure, PSL Research University, UPMC Univ. Paris 6,
 Sorbonne Universit\'es, Universit\'e Paris Diderot, Sorbonne Paris Cit\'e,   CNRS, 75005 Paris, France
}


\begin{abstract}
A shear flow of liquid metal (Galinstan) is driven in an annular channel by counter-rotating traveling magnetic fields imposed at the endcaps. When the traveling velocities are large, the flow is turbulent and its azimuthal component displays random reversals. Power spectra of the velocity field exhibit a $1/f^\alpha$ power law on several decades and are related to power-law probability distributions  $P(\tau)\sim \tau^{-\beta}$ of the waiting times between successive reversals. This $1/f$ type spectrum is observed only when the Reynolds number is large enough. In addition, the exponents $\alpha$ and $\beta$ are controlled by the symmetry of the system: a continuous transition between two different types of Flicker noise is observed as the equatorial symmetry of the flow is broken, in agreement with theoretical predictions.
\end{abstract}

\pacs{16 64}
                             
\maketitle


A puzzling problem in  physics is the ubiquity of '1/f' noise or 'Flicker' noise, i.e. the existence of a wide range of frequencies over which the low frequency power spectrum $S(f)$ of a physical quantity follows a power law $S(f)\sim f^{-\alpha}$, with $\alpha$ close to 1 (or more generally $0 < \alpha < 2$). Such behavior is observed in a broad variety of physical systems, ranging from voltage and current fluctuations in vacuum tubes or transistors~\cite{Hooge1981,Dutta1981} to astrophysical magnetic fields~\cite{Matthaeus1986}, and including biological systems~\cite{Gilden1995}, climate~\cite{Fraedrich2003} and turbulent flows~\cite{Dmitruk2007,Dmitruk2014, Ravelet2008,Costa2014}  to quote a few . 

Surprisingly, this ubiquity of $1/f$ noise does not seem to rely on a single explanation: although many interesting models have been proposed during the last 80 years, there is currently no universal mechanism for the generation of  $1/f$ fluctuations. 
Different levels of theoretical description of $1/f$ noise involve, the existence of a continuous distribution of relaxation times in the system~\cite{Bernamont1937,Vanderziel1950}, fractional Brownian motion ~\cite{Mandelbrot1968}, low dimensional dynamical systems close to transition to chaos ~\cite{Manneville1980,Procaccia1983,Geisel1987}. These systems often display an intermittent regime with bursts occurring after random waiting times $\tau$. For this type of point processes, it has been shown that a $f^{-\alpha}$ spectrum is related to a power law distribution $P(\tau) \propto \tau^{-\beta}$ with some relation between $\alpha$ and $\beta$ that depends on the symmetry of the signal~\cite{Lowen1993}. 

Although most of the early experimental observations of $1/f^{\alpha}$ noise do not display such discrete events in their time recordings, switching events have been observed in small electronic systems~\cite{Ralls1984} and more recently in blinking quantum dots~\cite{Kuno2000,Shimizu2001,Pelton2004}. These waiting times, distributed as a power-law, reflect the scale-free nature of the statistics, and are associated to durations spent by the system in two different states (bright or dark state in the case of quantum dots). More recently, statistical analysis of quasi-bidimensional turbulence of an electromagnetically forced flow exhibited a similar dynamics, in which a large scale circulation driven by a turbulent flow randomly reverses~\cite{Herault2015}. In this experiment, both $1/f$ power spectrum and power-law inter-event time probability distribution functions were observed. These results indicate that coherent structures generated in turbulent flows play a crucial role in the occurence of $1/f$ noise. On the other hand, it is known that such large scale coherent structures can exhibit very different dynamics depending on the level of turbulent fluctuations or the symmetry properties of the system. Whether these properties could affect $1/f$ noise is an open question. By carefully tuning the parameters of the experiment reported here, both the level of turbulence and the symmetry between two  states can be independently controlled, allowing for such investigation: we show how the occurence of $1/f$ fluctuations is directly related to the power-law PDF of waiting times, but critically depends on the level of turbulence generated in the flow. In addition, the  symmetry of the forcing plays a crucial role: different relations are satisfied by $\alpha$ and $\beta$ depending on whatever the two opposite states are symmetrical or not. In particular, a continuous transition between the different regimes predicted in~\cite{Lowen1993} can be obtained as a function of the skewness of the velocity PDFs, ultimately controlled by the symmetry of the external driving.\\

%
%

\begin{figure}[h]
\includegraphics[width=8 cm]{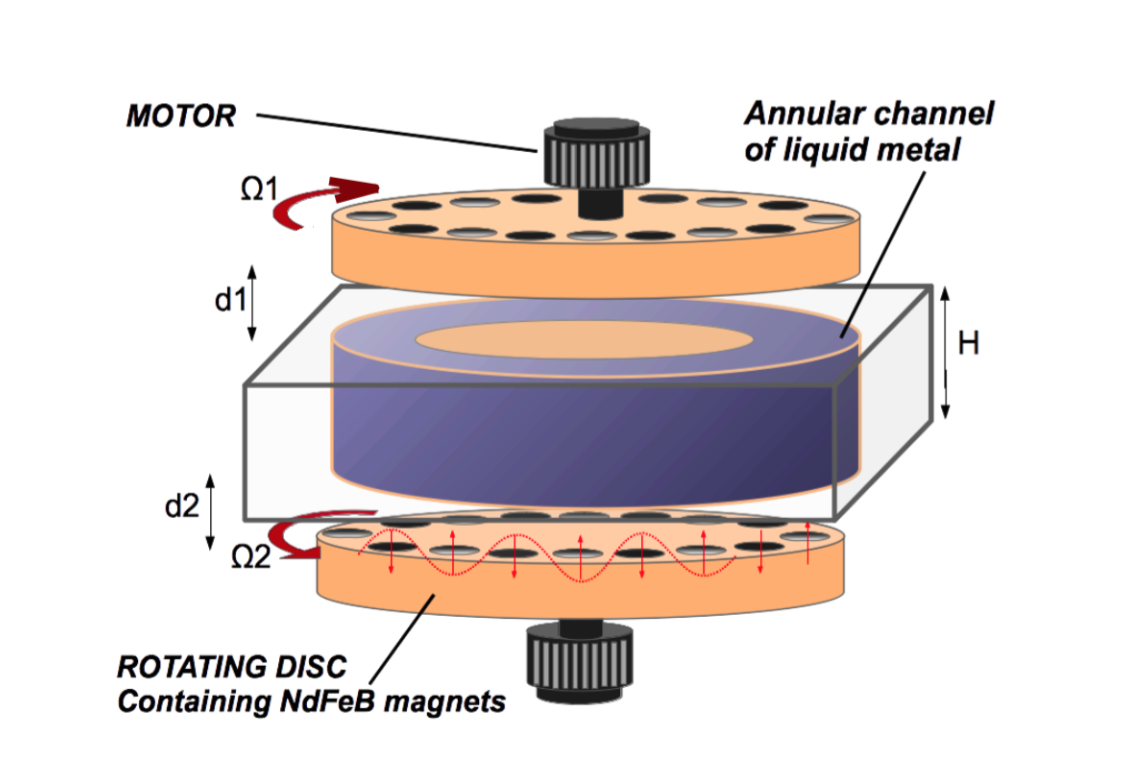}
\caption{Schematic view of the experiment. An annular channel made of PVC of mean radius $R=83$ mm and gap width $H=47$ mm is filled with a liquid metal (Galinstan).The flow is driven by the Lorentz force due to a traveling magnetic field (TMF) on each side of the experiment, created by $16$ Neodymium magnets placed on independently rotating discs.}
\label{setup}
\end{figure}

Fig.\ref{setup} shows a schematic picture of the experiment : an annular channel made of Polyvinyl Chloride (PVC), with inner radius $r_i=65$ mm, outer radius $r_o=98$ mm, and vertical height $H=47$ mm,  is filled with liquid Galinstan (GaInSn), an eutectic alloy which is liquid at ambiant temperature, with  kinematic viscosity $\nu=0,37 \cdot10^{-6}$ $m^{2}.s^{-1}$, density $\rho=6,44\cdot10^{3}$ $kg.m^{-3}$ and electrical conductivity $\sigma=3,46\cdot10^{6}$ $S.m^{-1}$.

At a distance $h=10$mm above and below the channel are located two rotating discs, each containing $16$ Neodymium magnets disposed with a regular spacing along a circle of radius $R=83$mm. These magnets are cylinders of diameter $d_m=20$mm and height $h_m=10$ mm, generating a magnetic field  $B_m^0=0.45T$ at their surface. They are arranged such that two adjacent magnets, separated by a distance $d_m=2\pi R/16=32.5$mm, are oriented with opposite polarity. The rotating discs therefore generate on each side of the channel a spatially periodic magnetic field traveling in the azimutal direction with an angular frequency $\Omega_i=2\pi f_i/16$ and a wavenumber $k=\pi/d_m$, where $f_i$ is the rotation frequency of the disc $i$ . The flow is  electromagnetically driven by the Lorentz force due to these traveling magnetic fields (TMF) and their related induced electrical currents. The frequencies of the discs $f_1$ and $f_2$ can be changed independently. This leads to the definition of $4$  dimensionless control parameters for the experiment: $F=(f_1-f_2)/(f_1+f_2)$ controls the asymmetry of the forcing provided by the top and bottom discs and $Re=[(f_1+f_2)/2]H^2/\nu$ is the Reynolds number based on the mean frequency of the discs. In addition, one can define the magnetic Prandtl number $Pm=\nu\mu_0\sigma$, which is of order $Pm\sim 10^{-6}$ for Galinstan, and the dimensionless magnetic field of the magnets which is represented by the Hartman number $Ha^2=B_0^2 \sigma H /(k \rho \nu)$, where $B_0$ is the magnetic field measured in the midplane of the channel and k is the wave number. For the experiments reported here, Ha=90. The velocity field is measured through Ultrasound Doppler Velocimetry (UDV) using three probes located in three different horizontal planes $z=0$ (midplane) and $z=\pm11 mm$. 

When the two traveling magnetic fields imposed at top and bottom endcaps rotate in the same direction, a strong azimuthal Lorentz force drives the flow in the same direction than the discs, and the device therefore acts as an induction pump. In that case, the velocity of the flow increases with both the magnitude of the applied field and the rotation rate of the discs. Note however that the fluid velocity is always smaller than the speed of the discs, and can be much smaller if the magnetic field is expelled outside the channel at large magnetic Reynolds number, $Rm=Re Pm$~\cite{Gissinger2016,Rodriguez2016}. 

\begin{figure}[h]
	\includegraphics[width=9.5cm]{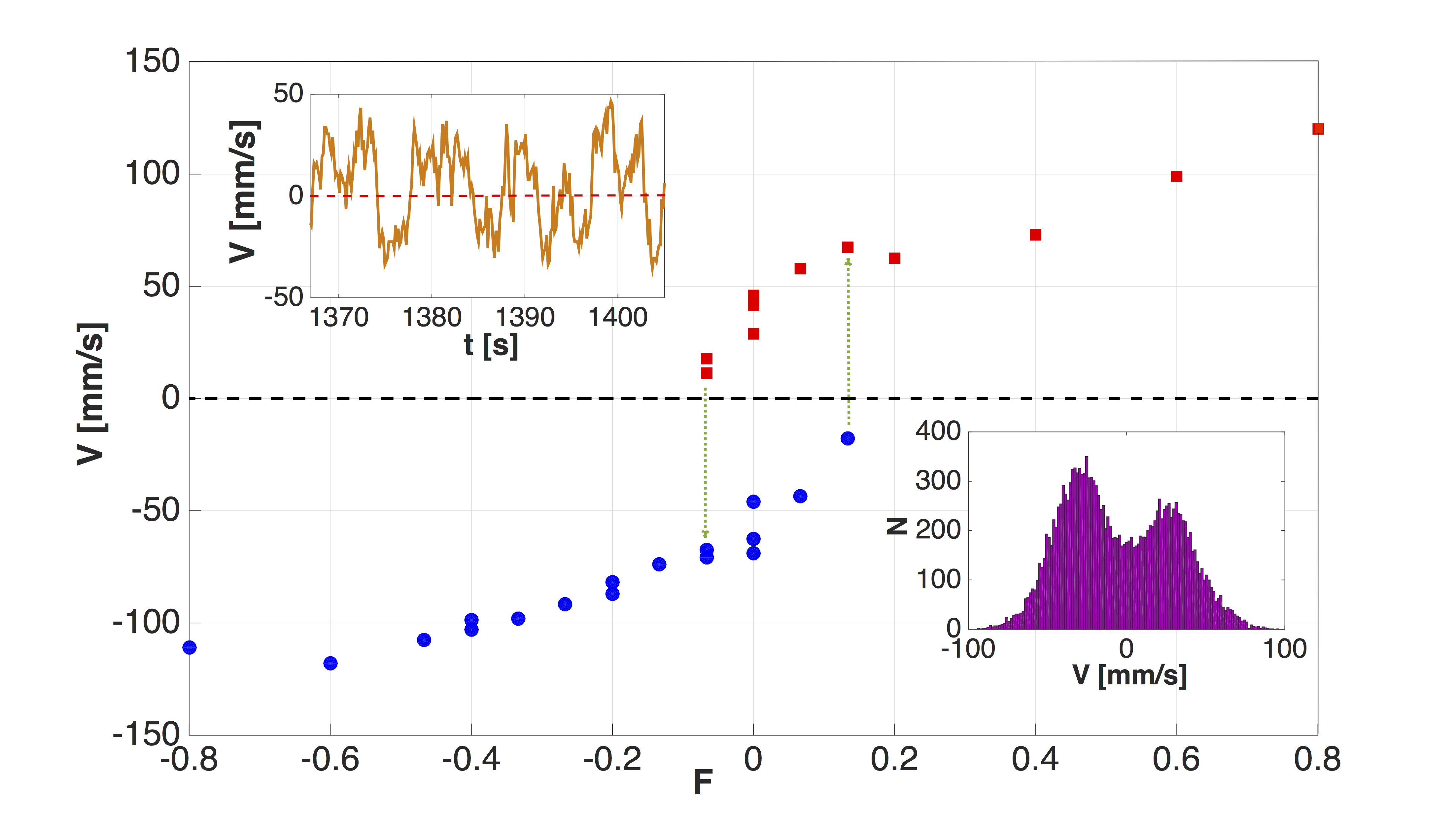}
	\caption{Most probable velocities measured in the midplane as a function of $F$, for $Re=7,1\cdot10^{3}$. The two vertical dashed lines indicate the region of bistability between positive and negative flow velocity. Upper-left inset: time series of the velocity in the bistable regime. Lower-right inset: bimodality of the PDF related to the bistability of the flow. }
		\label{cycle}
\end{figure}

We  focus here  on the configuration in which the two discs are counter-rotating. A strong shear flow develops in the channel, due to the opposing Lorentz forces generated at the top and bottom boundaries by the corresponding traveling magnetic fields. Fig. \ref{cycle} shows the bifurcation of the most probable velocities measured by UDV in the midplane of the channel,  as a function of the asymmetry parameter $F$. Red squares (respectively blue circles) indicate positive (resp. negative) mean velocity, meaning that the fluid in the midplane moves in the same direction than the upper (resp. lower) disc. Close to $F=0$, a bistability between this two states is observed. The upper-left inset shows a typical time series of the velocity field in this regime (here for F=0): the instantaneous velocity is strongly fluctuating and exhibits chaotic reversals of its polarity, the fluid following alternatively one disc or the other. As a consequence, the corresponding probability density function (PDF) shows a bimodal structure (lower-right inset), characterized by  two maxima in the PDF. The two vertical dotted lines delimitate the region for which such a bistability between positive and negative velocity is observed (characterized by bimodal PDFs). Note that the bifurcation diagram should be symmetrical with respect to $F=0$ exactly, for which none of the two states is favored by the forcing. In practice, the curve is slightly shifted to positive values (symmetrical PDFs obtained for $F\sim 0.05$ for this Reynolds number), which may be due to some imperfections in the experimental setup. Similar reversals of the velocity field have been described in von Karman swirling flows, in which the shear layer generated by two counter rotating bladed discs can undergo chaotic jumps from the midplane~\cite{Ravelet2008}. UDV measurements above and below the midplane indicate that a similar large scale dynamics of the central shear layer occurs in the present experiment.

\begin{figure}[h]
	\includegraphics[width=9cm]{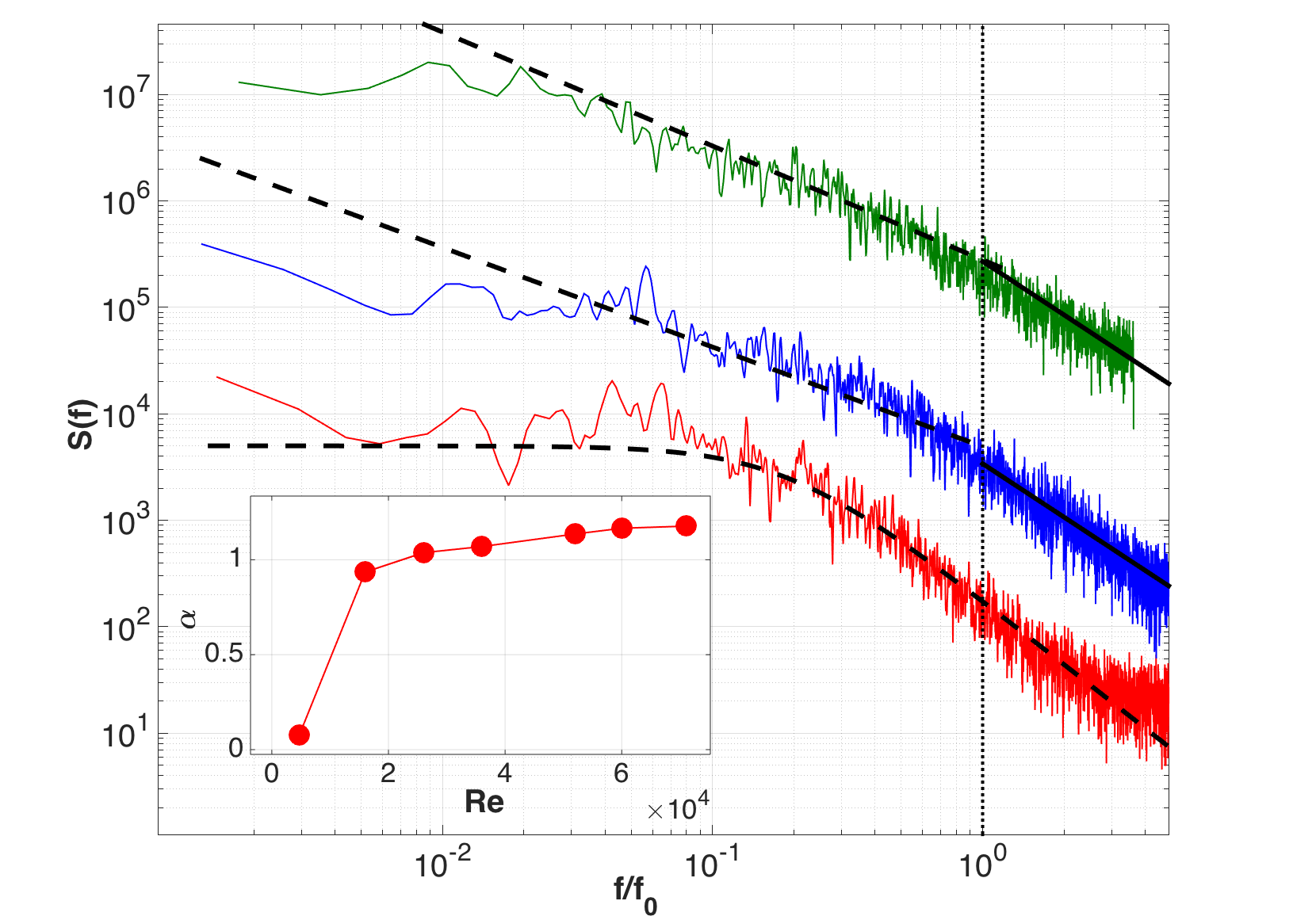}
	\caption{Frequency power spectra $S(f)$ of the velocity $V(t)$ for different Reynolds numbers ($Re=4,7\cdot 10^{3}$ ; $Re=1,6\cdot 10^{4}$  and  $Re=3,6\cdot 10^{4}$ from bottom to top). For clarity, the spectra have been multiplied by 1, 10 and 1000. Note the $-\frac{5}{3}$ slope at high frequency, and the occurence of $1/f^\alpha$ noise at low frequency. Inset: exponent $\alpha$ as a function of $Re$.}
		\label{spectre}
\end{figure}

We first study the evolution of the statistical properties of the velocity field in the bistable regime, for $F\sim 0$. In Fig.\ref{spectre}, we report the frequency power spectra extracted from time series $V(t)$ measured by UDV in the midplane, for different values of the Reynolds number. First  note that all  power spectra show a  $f^{-\frac{5}{3}}$ direct cascade of energy from the injection scale $f_0\sim \frac{U}{H}$, where $U$ is the mean velocity of the flow (measured close to each disc) and $d$ is the gap of the channel. We focus here on the behavior of the spectra at frequency below the injection scale $f_0$. We first observe that they strongly depend on the Reynolds number: at the lowest $Re$, the spectrum is flat for $f \ll f_0$, but as $Re$ is increased beyond a critical value $Re_c\sim 10^4$, the system shows a build up of the energy towards low frequency, such that $1/f^{\alpha}$  noise is observed for large Reynolds numbers. The inset of Fig.\ref{spectre} shows the dependence of $\alpha$ on $Re$, and suggests that it rapidly converges to values slightly larger than $\alpha=1$  in the limit of  large $Re$. We emphasize that the spectra below the injection scale $f_0$ are not related to any turbulent cascade process since the frequencies are too low to correspond to any spatial scale within the fluid container. In particular, the $1/f$ spectra observed here in the bulk flow are not similar to the $1/f$ spectra observed  in turbulent boundary layers that trace back to $1/k$ spectra through the Taylor hypothesis~\cite{Perry1986}.

\begin{figure}[h]
	\includegraphics[width=8cm]{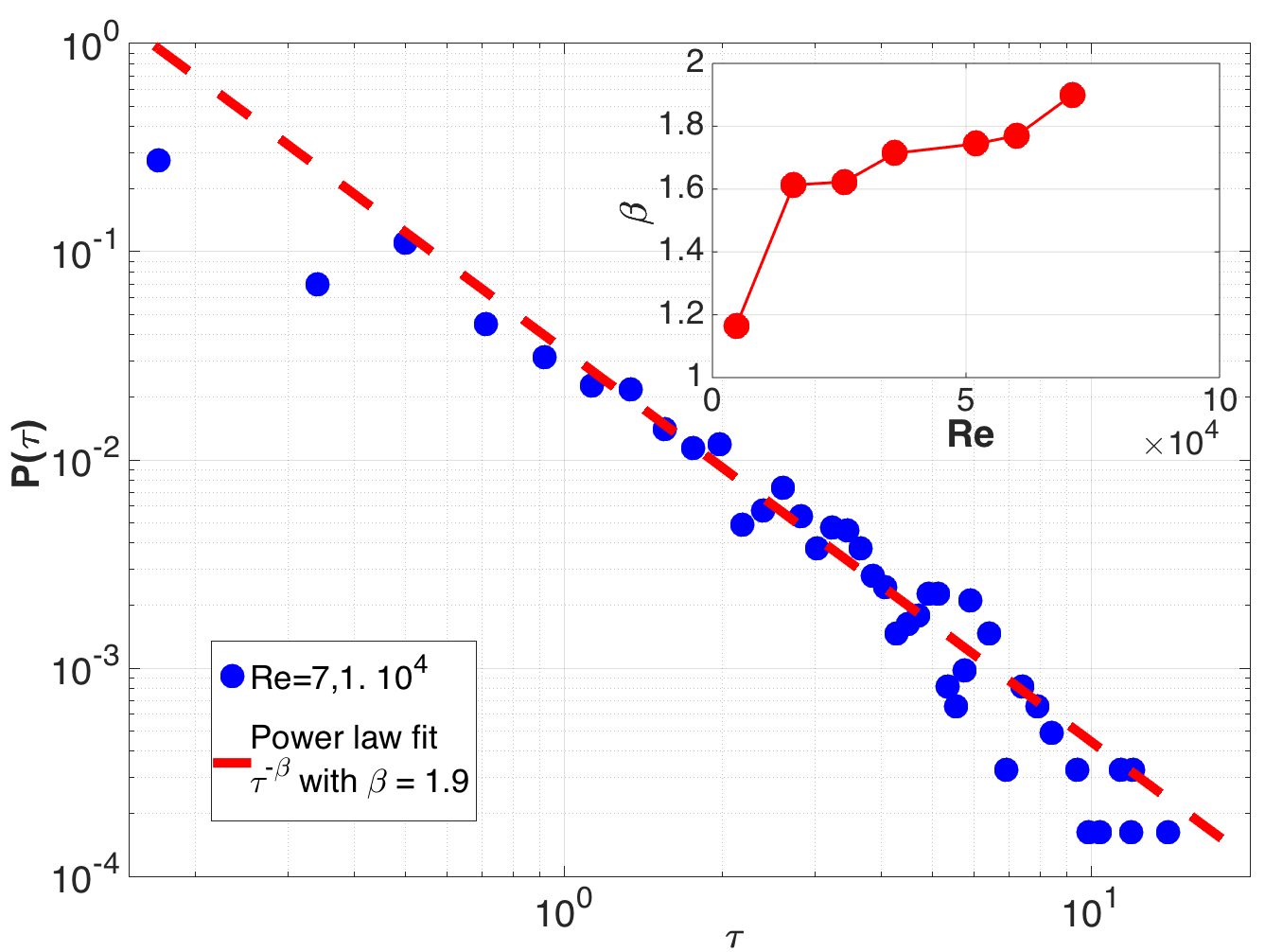}
	\caption{Distribution of the waiting time $P(\tau)$ between two successive reversals of the flow for $Re=7,1\cdot 10^{4}$. For sufficiently large $Re$, the distribution follows a power law $\tau^{-\beta}$. Inset: $\beta$ as a function of $Re$.}
	\label{dist}
\end{figure}

Since these results have been obtained for $F\sim 0$, all the power spectra shown in Fig.\ref{spectre} are related  to time series exhibiting chaotic reversals between two symmetrical states. These random reversals can be characterized by the distribution $P(\tau)$ of the waiting time (WT) $\tau$ between two successive transitions, as shown in Fig.~\ref{dist} for $Re=7,1\cdot10^4$. We observe  that the waiting times are distributed according to a power law $P(\tau)\sim \tau^{-\beta}$, in contrast to the  exponential distribution generally observed in the case of a memoryless system. The presence of such power-law PDF therefore suggests a more complex non-Poissonian physics underlying the occurence of polarity changes.  Note that similarly to $\alpha$, the exponent of the power law depends on $Re$, and slowly tends to $\beta=2$ as $Re$ is increased to large values.

\begin{figure}[h!]
	\includegraphics[width=9cm]{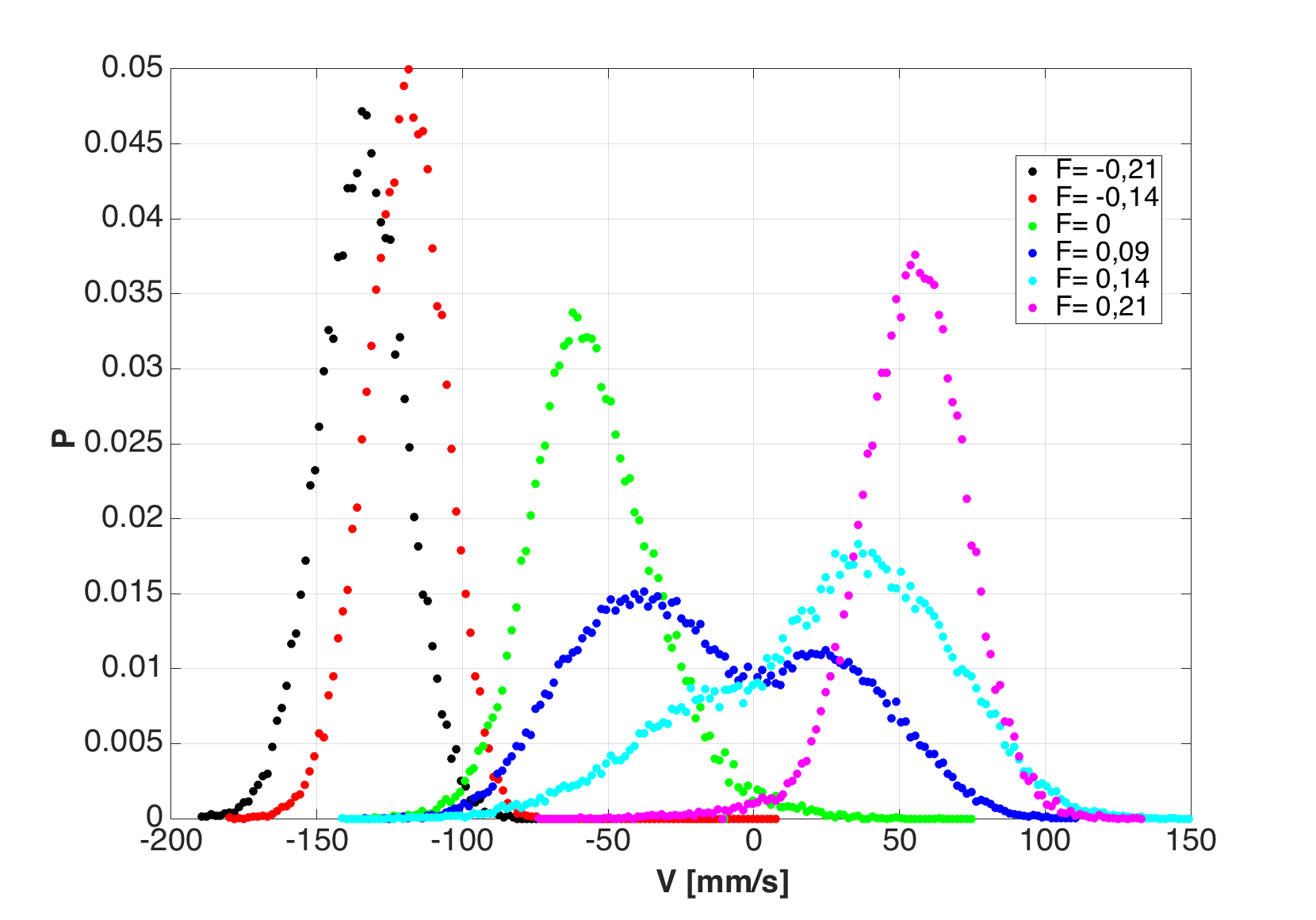}
	\caption{Probability density function of the velocity field for different values of the asymmetry  $F$ of the forcing. Note the transition from a Gaussian distribution at large $|F|$ to bimodal behavior for $F\sim 0$.}
	\label{PDF_F}
\end{figure}
\begin{figure}[h!]
	\includegraphics[width=9cm]{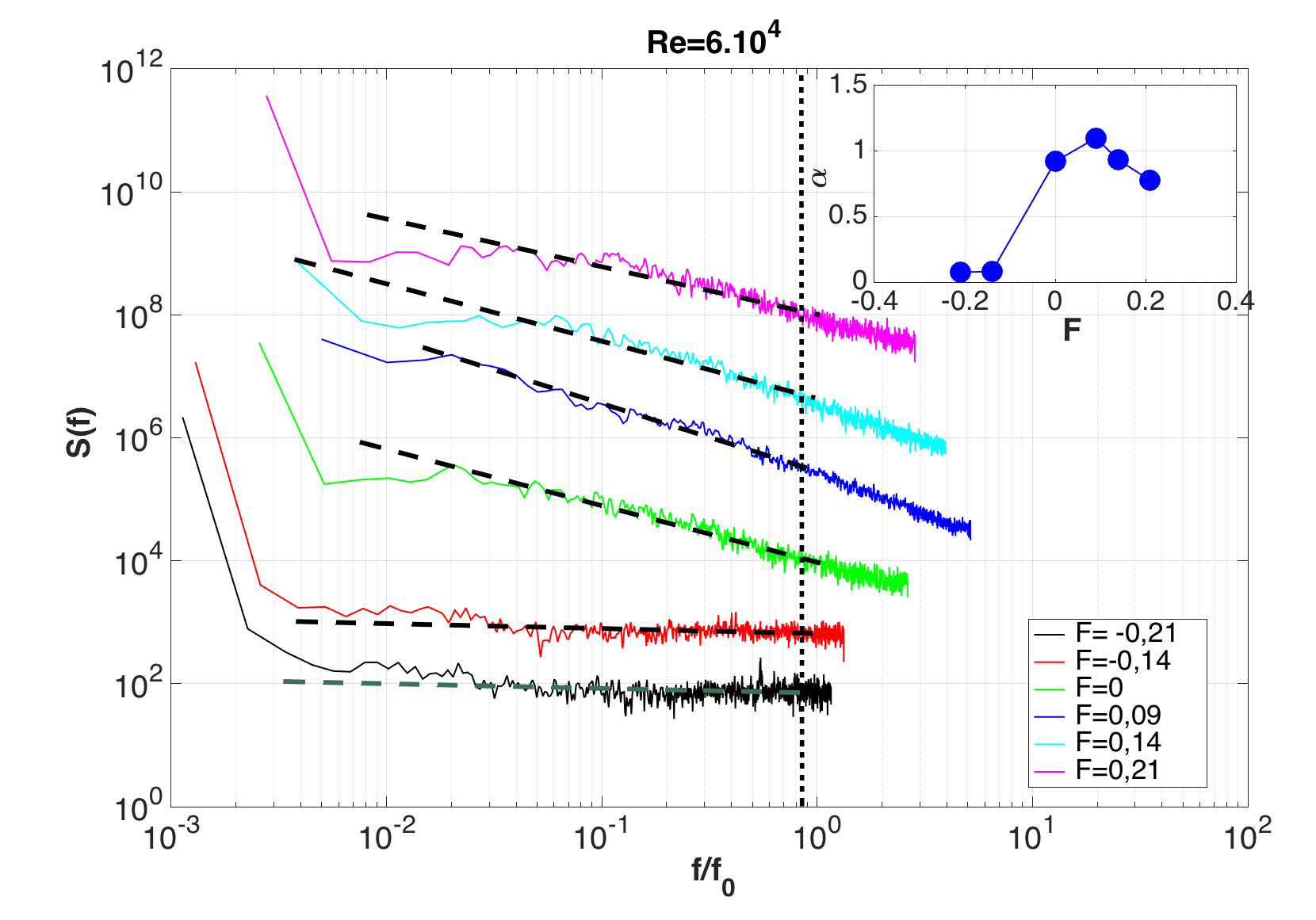}
	\caption{: Frequency power spectra $S(f)$ of the velocity $V(t)$ for different values of the asymmetry parameter $F$. For clarity, the spectra have been multiplied by $10$ for each increment of $F$. Inset shows the exponent $\alpha$ as a function of $F$.}
	\label{spec_F}
\end{figure}

The exponents of the power spectra and of the WT distribution also strongly depend on the asymmetry of the magnetic forcing, controlled by the value of $F$. In Fig. \ref{PDF_F}, we report the probability density function (PDF) of the velocity field in the midplane, for various values of F and a fixed value of the Reynolds number $Re=6.10^4$. When $F$ has large negative or positive values, the system is in a non-reversing regime with negative (respectively positive) mean velocity, and the fluid follows the bottom (respectively the top) disc with a gaussian distribution of the velocity fluctuations. For values of $F$ close to $0$, the distribution is either  bimodal and roughly symmetrical with respect to zero (for instance $F=0.09$),  or asymmetrical with a non-gaussian tail (for instance $F=0.14$). 

Interestingly, this asymmetry in the forcing clearly controls the value of the exponent of the power spectrum at low frequency, as shown by Fig.\ref{spec_F}. For strongly asymmetric forcing ($F=-0.21$ and $F=-0.14$), the spectrum is flat, with $f^0$ behavior on several decades for $f<f_0$. As the flows starts to randomly explore the other polarity, $\alpha$  increases, even when the corresponding PDF is not bimodal (see $F=0$). The exponent $\alpha$ reaches its maximum value $\alpha=1.1$ for symmetrical PDFs of the velocity, and then decreases again with $F$ as the flows come back to a non-reversing state.

In fact, it has been shown~\cite{Lowen1993,Niemann2013} that in the presence of a heavy-tailed distribution similar to the one shown in Fig.\ref{dist}, the exponent $\alpha$ of the power spectrum and the exponent $\beta$ of the WT distribution are related: in the case of a symmetric process (meaning that the two states have similar transition probability), one expects the relation $\alpha+\beta=3$, whereas $\beta-\alpha=1$ is predicted in a non-symmetric process (e.g. for random bursts). It has been shown in~\cite{Herault2015b} that one prediction or the other can be observed in different experiments : for instance, pressure fluctuations in 3D turbulence~\cite{Abry1994} follow $\beta-\alpha=1$ scaling, whereas $\alpha+\beta=3$  is observed for random reversals of a large scale flow generated by Kolmogorov forcing~\cite{Herault2015b}.

\begin{figure}[h!]
	\includegraphics[width=7.5cm]{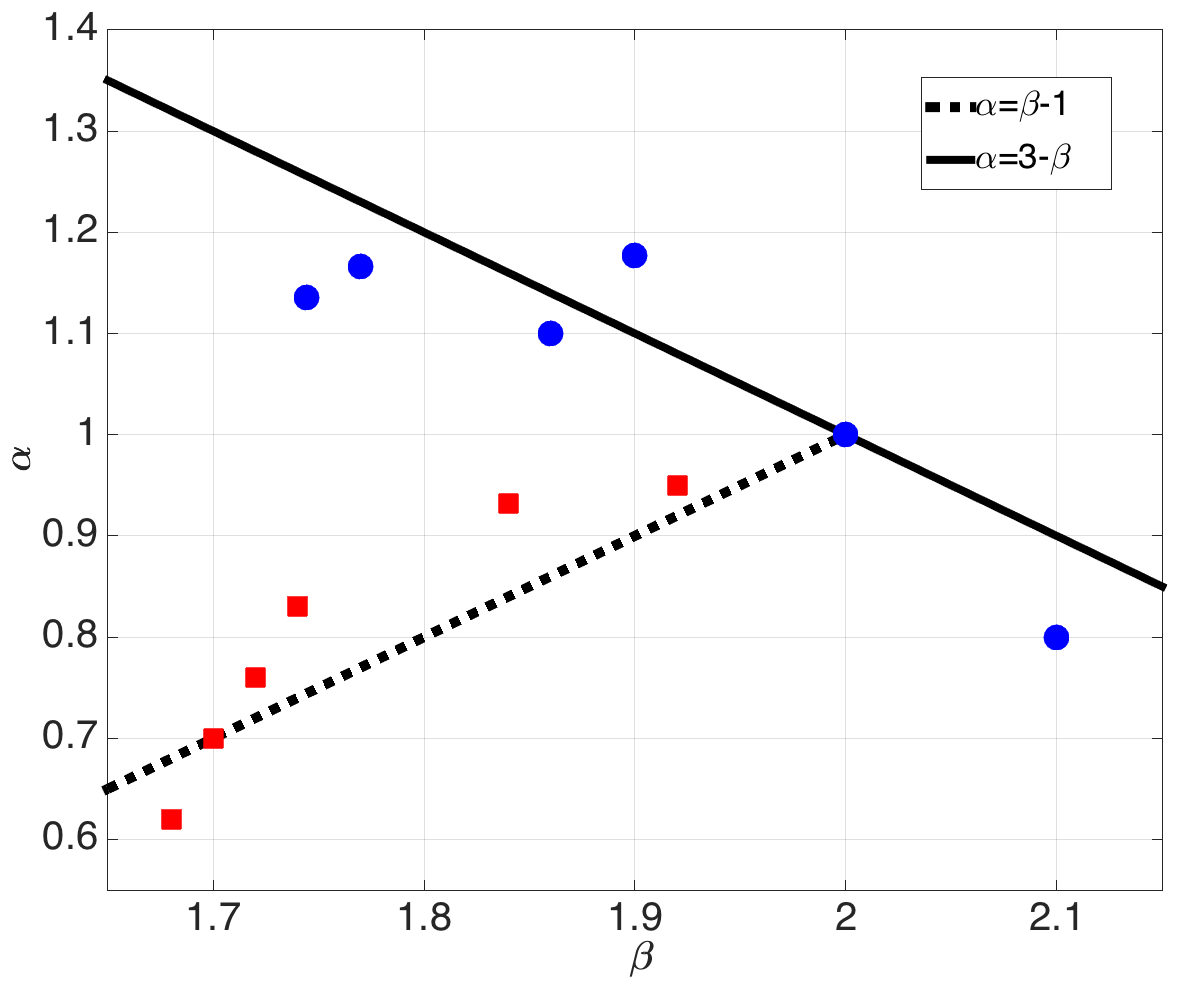}
	\caption{$\alpha$ as a function of $\beta$ for various values of $F$ and $Re$. The dashed line indicates the regime $\beta-\alpha=1$, while the solid line corresponds to $\beta+\alpha=3$. The skewness $\theta$ of the PDFs determines in which regime the system lies ($\theta <0.1$ for blue circles, $\theta>0.1$ for red squares).}
	\label{exponents}
\end{figure}

We show here that both regimes can be observed in the same experiment and for the same measured quantity, depending only on the asymmetry parameter $F$ and the Reynolds number: Fig~\ref{exponents}  reports most of our experimental runs (obtained for various values of $F$ and $Re$) in the parameter space $\{\alpha,\beta\}$, in which the dashed line indicates the regime $\beta-\alpha=1$ and the solid line indicates $\alpha+\beta=3$. For each point, we have computed the skewness of the PDFs of the velocity $\theta=\langle [(V(t)-\mu)/\sigma)]^3 \rangle$, where $\sigma$ and $\mu$ are respectively the standard deviation and the mean.
When the probability density function of the flow exhibits a roughtly bimodal distribution ($\theta<0.1$, blue circles), most of the points tend to collapse on the line $\alpha+\beta=3$, while asymmetrical reversals ($\theta>0.1$, red squares) lie along $\beta-\alpha=1$, valid for bursting processes only.  While these results show that the asymmetry of the forcing controls the type of 1/f noise (i.e. the value of the sum or the difference of the exponents) which is observed, what controls exactly the values of the exponents remains unclear.

It is also important to note that Fig.\ref{exponents} reports results obtained only for sufficiently large Reynolds numbers (in practice $Re\ge5\cdot 10^4$) and $F$ not too large (keeping only non-gaussian distributions). 

The problem we studied experimentally is related to the general question of the low frequency behavior of the turbulent velocity spectrum. As seen in Fig. \ref{spectre}, the power increases at low frequency as the Reynolds number is increased. This could be somewhat surprising since the phenomenology of three-dimensional turbulence predicts an increase of the inertial range toward the small spatial scales. The increase of power at low frequency results from the instability of the shear layer that develops on the turbulent background. We therefore showed that the low frequency behavior of turbulent flows is strongly related to the dynamics of coherent structures, here the shear layer, and confirmed observations made in several other flow configurations~\cite{Herault2015b}. As shown in another context, large scale instabilities of turbulent flows can be modeled by keeping only large scale modes that obey the truncated Euler equation (TEE)~\cite{Shukla2016}. Numerical simulations of  TEE have displayed $1/f$ spectra~\cite{Shukla2018}. Numerical simulations of this type of models are presently studied in the case of a turbulent shear layer. We emphasize that this process for generating 1/f noise involves a large number of degrees of freedom with many triads in nonlinear interaction and therefore strongly differs from low dimensional dissipative dynamical systems. 
Our experimental results also show that although the power law exponents $\alpha$ and $\beta$ only slightly  change with $Re$, they strongly depends on the asymmetry parameter. This second observation is interesting, because the continuous transition from $\alpha+\beta=3$ to $\beta-\alpha=1$ generated as the equatorial symmetry of the flow is broken shows that both regimes can be observed within the same system. In other words, some features of $1/f$ noise can be directly related to the asymmetry of the system. It would be interesting to see if this relation can be used to understand some systems from the characteristics of their $1/f$ fluctuations. For instance, the study of the exponents $\alpha$ and $\beta$ from $1/f$  fluctuations of the solar wind or the luminosity of some stars may help probing their symmetry properties, less accessible from observations. 
From a fundamental viewpoint, it could be argued that we have replaced the problem of finding a mechanism for $1/f$ noise by the problem of providing an explanation for the power law PDF of waiting times. However, this could be a useful step since a generic mechanism has been proposed for the later~\cite{Montroll1982}. Finally, we can understand the particular role played by the value $\alpha = 1$ that is common to the symmetric and asymmetric cases (see Fig.\ref{exponents}). With symmetric forcing, we expect  $\alpha = 1$ to be selected by small asymmetric perturbations.

\bibliography{biblio}

\begin{thebibliography}{} 



\bibitem{Hooge1981}
F. N. Hooge, T. G. M. Kleinpenning and L. K. J. Vandamme, Rep. Prog. Phys. {\bf 44}, 479 (1981)

\bibitem{Dutta1981}
P. Dutta and P.M. Horn, Rev. Mod. Phys. {\bf 53}, 497 (1981)


\bibitem{Matthaeus1986}
W.H. Matthaeus and M. L. Goldstein, Phys. Rev. Lett. {\bf 57}, 495 (1986)

\bibitem{Gilden1995}
D.L. Gilden, T. Thornton, M.W. Mallon,Science, {\bf 267}, 1837-1839 (1995). 

\bibitem{Fraedrich2003}
K. Fraedrich and R. Blender, Phys. Rev. Lett. {\bf 90}, 108501 (2003)

\bibitem{Dmitruk2007}
P. Dmitruk and W.H. Matthaeus, Phys.Rev.E {\bf 76}, 036305 (2007).

\bibitem{Dmitruk2014}
P. Dmitruk et al, Phys.Rev.E {\bf 90}, 043010 (2014).


\bibitem{Ravelet2008}
F. Ravelet, A. Chiffaudel and F. Daviaud, J. Fluid. Mech. {\bf 339}, 601 (2008)

\bibitem{Costa2014}
A. Costa et al., Phys. Rev. Lett. {\bf 113}, 108501 (2014)

\bibitem{Bernamont1937}
J. Bernamont, Proc. Physical Soc. {\bf 49}, 138 (1937)
  
\bibitem{Vanderziel1950}
A. van der Ziel, Physica {\bf 16}, 359 (1950)

\bibitem{Mandelbrot1968}
B. B. Mandelbrot and I. W. van Ness, SIAM {\bf 10}, 422 (1968)

\bibitem{Manneville1980}
P. Manneville, J. Physique {\bf 41} 1235 (1980).

\bibitem{Procaccia1983}
I. Procaccia and H. Schuster, Phys. Rev. A  {\bf 28}, 1210 (1983)

\bibitem{Geisel1987}
T. Geisel, A. Zacherl and G. Radons, Phys. Rev. Lett. {\bf 59}, 2503 (1987)

\bibitem{Lowen1993}
S.B. Lowen and M. Teich. C., Phys. Rev. E, {\bf 47} , 992 (1993)

\bibitem{Ralls1984}
K. S. Ralls K. S. et al., Phys. Rev. Lett. {\bf 52}, 228 (1984)

\bibitem{Kuno2000}
M. Kuno et al, J. Chem. Phys. {\bf 112}, 3117 (2000)

\bibitem{Shimizu2001}
K. T. Shimizu et al., Phys. Rev. B {\bf 63}, 205316 (2001)

\bibitem{Pelton2004}
M. Pelton, D. G. Grier, and P. Guyot-Sionnest,  Appl. Phys. Lett. {\bf 85} 819 (2004).

\bibitem{Herault2015}
J. Herault, F. Petrelis, S. Fauve,  Euro. Phys. Lett. {\bf 111} 44002 (2015).

\bibitem{Gissinger2016}
C. Gissinger, P. Rodriguez-Imazio, S. Fauve, Phys. Fluids. {\bf 28} 034101 (2016)

\bibitem{Rodriguez2016}
P. Rodriguez-Imazio and C. Gissinger, Phys. Fluids. {\bf 28} 034102 (2016)

\bibitem{Perry1986} 
A. E. Perry, S. Henbest and M. S. Chong, J. Fluid Mech. {\bf 165}, 163 (1986) and references therein.

\bibitem{Niemann2013}
M. Niemann, H. Kantz and E. Barakai, Phys. Rev. Lett., {\bf 110}, 140603 (2013)

\bibitem{Herault2015b}
J. Herault, F. Petrelis, S. Fauve,  Journ. Stat. Phys. {\bf 161} 1379 (2015).

\bibitem{Abry1994}
P. Abry et al, J. Physique II, {\bf 4}, 725 (1994) 

\bibitem{Shukla2016}
V. Shukla, S. Fauve and M. Brachet, Phys. Rev. E {\bf 94}, 061101 (2016).

\bibitem{Shukla2018} V. Shukla, S. Fauve and M. Brachet, 1/f noise in Kolmogorov flows, in preparation (2018).

\bibitem{Montroll1982}
E. W. Montroll and M. F. Shlesinger, Proc. Natl. Acad. Sci. USA {\bf 79}, 3380 (1982)




\end{thebibliography}

\end{document}